\documentclass[amssymb,amsmath,aps,showpacs,floatfix,nofootinbib,showpacs,12pt]{revtex4}
\usepackage{mathrsfs}
\usepackage{amssymb}
\usepackage{graphicx}
\usepackage{color}
\usepackage{soul}
\usepackage{latexsym}

\newcommand{\lsim}{\lesssim}
\newcommand{\gsim}{\gtrsim}

\def\lsim{\mathrel{\raise.3ex\hbox{$<$\kern-.75em\lower1ex\hbox{$\sim$}}}}
\def\gsim{\mathrel{\raise.3ex\hbox{$>$\kern-.75em\lower1ex\hbox{$\sim$}}}}

\def\beq{\begin{equation}}
\def\eeq{\end{equation}}
\def\beqn{\begin{eqnarray}}
\def\eeqn{\end{eqnarray}}
\def\bea{\begin{eqnarray}}
\def\eea{\end{eqnarray}}
\def\be{\begin{equation}}
\def\ee{\end{equation}}
\newcommand{\fslash}[1]{{#1 \kern -0.7em/ \kern 0.1em}}

\begin{document}

\voffset 1.25cm

\title{Prospects for Detecting Neutrino Signals from Annihilating/Decaying
Dark Matter to Account for the PAMELA
and ATIC results}

\author{Jia Liu, Peng-fei Yin and Shou-hua Zhu }

\affiliation{Institute of Theoretical Physics $\&$ State Key
Laboratory of Nuclear Physics and Technology, Peking University,
Beijing 100871, P.R. China}
\date{\today}

\begin{abstract}

Recent PAMELA data show that positron fraction has an excess above
several GeV while anti-proton one is not. Moreover ATIC data
indicates that electron/positron flux have a bump from 300 GeV to
800 GeV. Both annihilating dark matter (DM) with large boost factor
and decaying DM with the life around $ 10^{26} s$ can account for
the PAMELA and ATIC observations if their main final products are
charged leptons ($e$, $\mu$ and $\tau$). In this work, we calculated
the neutrino flux arising from $\mu$ and $\tau$ which originate from
annihilating/decaying DM, and estimated the final muon rate in the
neutrino telescopes, namely Antares and IceCube. Given the excellent
angular resolution, Antares and IceCube are promising to discover
the neutrino signals from Galactic center and/or large DM subhalo in
annihilating DM scenario, but very challenging in decaying DM
scenario.

\end{abstract}


\maketitle

\newpage

\section{Introduction}

The main components of our Universe are `dark'. Many astrophysical
observations have confirmed the existence of dark matter (DM) which
contributes roughly $23\%$ to the energy density of the Universe. If
the DM are weakly interacting massive particles (WIMPs), they would
annihilate into gamma rays, neutrinos and anti-matter particles,
which could be detected experimentally. Generally speaking, the DM
should be stable compared with the life of the Universe. However it
is possible that the DM is decaying at an extremely low rate.
Similarly, the DM decay products can be observable.

Recently PAMELA released their first cosmic rays results on the
positron and anti-proton ratios \cite{Adriani:2008zr}. Usually the
anti-matter particles are expected to be produced when the cosmic
rays propagate in the Galaxy and interact with the interstellar
medium. The PAMELA data show a clear upturn (excess) on the positron
ratio above $\sim 10 $ GeV up to $\sim 100 $ GeV. Besides PAMELA
observation, ATIC collaboration also reported an electron/positron
excess around $300 \sim 800$ GeV \cite{Chang:2008zz}. Contrary to
the excess of position ratio, PAMEMA anti-proton ratio is consistent
with the predictions. Both PAMELA and ATIC observations indicate
some unknown high energy primary positron sources in the Galaxy.
Studies showed that the pulsars or other nearby astrophysical
sources may account for the PAMELA results \cite{pulsar} .

Other possibility to induce the positron excess can be the
annihilating \cite{annidm,Cirelli:2008jk,ArkaniHamed:2008qn} or
decaying \cite{decaydm} DM in the Galaxy. In the annihilating DM
scenario, if the cross section $ \left\langle {\sigma v}
\right\rangle$ is taken as the typical annihilation cross section
$\sim 3\times10^{-26}cm^3s^{-1}$ for thermal relic, a large extra
boost factor (BF) is needed to produce sufficient positron flux to
fit the PAMELA results. Many ideas are proposed to induce the large
BF. For example, the new `long' range force would enhance the
annihilation cross section for today low velocity DM, which is
dubbed as `Sommerfeld effect'
\cite{Cirelli:2008jk,ArkaniHamed:2008qn,sommerfeld,Lattanzi:2008qa}.
For the decaying DM scenario, by adjusting the lifetime of the DM,
no BF is needed. Contrary to positron excess, no obvious anti-proton
excess has been observed. This suggests that the main final products
from the annihilation or decay of DM are charge leptons ($e$, $\mu$
and $\tau$) for $m_{DM}< 1TeV$. Otherwise one needs much heavy DM,
say $O(10TeV)$ \cite{Cirelli:2008jk}, or some special cosmic-ray
propagation parameters.

Provided that charged leptons are generated by DM annihilating
and/or decaying DM, it is quite interesting to investigate which
charged leptons are response for the PAMELA and ATIC observations.
It is natural to expect that the high energy charged leptons are
correlated to other observable signals, such as gamma rays,
synchrotron emissions or neutrinos. Detecting neutrino signals from
DM is very attractive and useful to pin down the nature of DM.
In this paper, we will calculate the neutrino flux from the $\mu$
and $\tau$ which are producing by DM. Unlike other DM related
products, neutrinos have less trajectory defection and less but
finite energy loss \cite{itoh} during propagation, due to their
weakly interaction nature. As a result, the neutrinos can keep the
important information of the DM. For the same reason, the neutrinos
are difficult to be detected if the flux is low (for some recent
works, see
\cite{Beacom:2006tt,Barger:2007xf,PalomaresRuiz:2007ry,Liu:2008kz,Yin:2008mv,Covi:2008jy}
). Usually the neutrinos detected by neutrino telescopes, such as
Super-Kamiokande \cite{Desai:2004pq,Ashie:2005ik}, AMANDA
\cite{Ackermann:2004aga} etc. are expected to be the atmospheric
neutrinos, which are produced by interaction between the cosmic rays
and atmosphere. The atmospheric neutrinos are the main backgrounds
for detecting the extra neutrinos from DM. Currently, no high energy
astrophysical neutrinos have ever been detected
\cite{Kistler:2006hp}.

If the charged leptons are indeed produced by DM, it is a good news
for neutrino detection. Firstly, the neutrino flux can be much
larger than that of usual expected. The PAMELA and ATIC data need
higher primary positron production rate than usual production, thus
it is natural to expect that large neutrino flux can be produced
from $\mu$ and $\tau$ decays. Secondly, the heavy DM (e.g. $\sim
TeV$) is favored by ATIC result, thus the neutrinos at the similar
energy scale are expected. As the main backgrounds, the atmospheric
neutrino flux decreases logarithmically with the increment of
energy, thus higher energy neutrinos correspond to less backgrounds.
On-going and future kilometer size neutrino telescopes such as
Antares \cite{Aslanides:1999vq}, IceCube \cite{Ahrens:2003ix} and
KM3NeT \cite{Carr:2007zc} can better explore these signals. Thirdly,
in the annihilating DM scenario, the flux of neutrinos depends on
the square of the DM number density. If one searches the Galactic
center (GC) or large DM subhalos which contain the denser DM than
other areas, large neutrino flux can be produced. Moreover if the
Sommerfeld effect of DM annihilation is true, the DM annihilation in
the subhalo can acquire even larger BF due to the lower velocity
dispersion than that in the DM halo
\cite{ArkaniHamed:2008qn,Lattanzi:2008qa}. As a side remark, this
feature provides a potential method to distinguish two DM scenarios,
namely GC or DM subhalos can be higher flux neutrino sources in the
annihilating DM scenario than those in the decaying DM one.

This paper is organized as following. In section II, we calculated
the neutrino flux from the GC and DM subhalo, in the
model-independent manner, in the annihilating and decaying DM
scenarios. In section III, we estimated the muon rate at Antares and
IceCube for the two DM scenarios. We pointed out that Antares and
IceCube might detect neutrino signals in annihilating DM scenario
and distinguish different DM scenarios. Section IV contains our
discussions and conclusions.

\section{Neutrino signals from annihilating/decaying DM}

PAMELA data show an excess on positron fraction about $10 \sim 1000$
times larger than usual estimation, so it is natural to expect that
the neutrinos from DM are also enhanced by the similar number. In
the annihilating DM scenario, it is achieved by Sommerfeld
enhancement and/or clumps in DM density. In this paper, BF is used
to denote the Sommerfeld enhancement. Astrophysical enhancement is
contained in the astrophysical factor, as described below.

The neutrino flux observed on the Earth can be written as
\begin{equation}
\phi^{A}
(E,\theta)=\rho_{\odot}^2R_{\odot}\times\frac{1}{4\pi}\frac{\langle
\sigma v\rangle }{2m_{\chi}^2}\frac{{\rm d}N}{{\rm d}E}\times
J^{A}(\theta ) \label{ADMflux}
\end{equation}
and
\begin{equation}
\phi ^D (E,\theta ) = \rho _ \odot  R_ \odot   \times \frac{1}{{4\pi
}}\frac{1}{{m_\chi  \tau _\chi  }}\frac{{dN}}{{dE}} \times J^D
(\theta ), \label{DDMflux}
\end{equation}
with dimensionless $J^{A}(\theta )$ and $J^D (\theta )$ defined as
\begin{equation} J^{A}(\theta ) = \frac{1}{{\rho _ \odot ^2 R_ \odot }}\int_{LOS}
{\rho ^2 (l)dl} \label{ja}\end{equation}
\begin{equation}
 J^D (\theta ) =
\frac{1}{{\rho _ \odot R_ \odot }}\int_{LOS} {\rho (l)dl}
\label{jd}.\end{equation}
 Here 'A' and 'D' denote annihilating and
decaying DM respectively. $\rho_{\odot}=0.34$ GeV cm$^{-3}$ is the
local DM density and $R_{\odot}=8.5$ kpc is the distance between the
Sun and the GC. $\theta$ is defined as the angle between the
observational and the GC directions, $m_{\chi}$ is the mass of DM
particle, and ${\rm d}N/{\rm d}E$ is the energy spectrum of  $\nu$
per annihilation or decay. $\langle\sigma v\rangle$ is the thermally
averaged annihilation cross section, which can be written as
\begin{equation} \left\langle {\sigma v} \right\rangle  =
\left\langle {\sigma v} \right\rangle _0 \times BF.\end{equation}
Here $ \left\langle {\sigma v} \right\rangle _0$ is set to be $ 3
\times 10^{ - 26} cm^3 s^{ - 1}$, which is the typical annihilation
cross section for the present dark matter abundance under the
standard thermal relic scenario. Thus the number BF is enhancement
factor compared with the typical value. The BF can arise from the
Sommerfeld enhancement as described above. The integral path in Eqs.
(\ref{ja}) and (\ref{jd}) is along the line-of-sight (LOS). The
neutrino spectrum ${\rm d}N/{\rm d}E$ comes from $ \mu ,\tau$ decay.
We count all three flavor of neutrinos and anti-neutrinos at the
decay point. After vacuum oscillation, we adopt assumption that the
three flavor neutrinos have the equal flux \cite{Beacom:2006tt}.
Thus we multiply a factor of $ 1/3$ in order to  get the flux of
muon neutrino and anti-muon neutrino, which is the main measured
component in the neutrino detector. We should mention here that the
tau neutrinos can also be observed by the Cherenkov neutrino
detector because tau neutrino has the different signal compared with
muon neutrino. The tau neutrino can interact with nucleons and
produce tau leptons which subsequently decay into muons. Thus they
will produce a kink in the reconstruction by Cherenkov light, while
the muon neutrino has no such kink. The authors of Ref.
\cite{Covi:2008jy} have discussed the tau neutrino from decaying
dark matter and found it can improve the detection by a few orders
of magnitude. The reason is simply that the atmospheric neutrinos
has much less tau neutrinos than muon neutrinos at high energy. In
this paper, we will concentrate on muon neutrinos observation.

The solid angle average of this J factor is defined as
\begin{equation} J^{A,D}_{\Delta \Omega } = \frac{1}{{\Delta \Omega
}}\int_{\Delta \Omega } {J^{A,D}(\theta )d\Omega } \label{jadomega}
\end{equation} with $ \Delta \Omega = 2\pi (1 - \cos \theta )$.
The averaged neutrino flux in a certain cone can be written as
\begin{equation}
\phi _{\Delta \Omega }^A (E) = \rho _ \odot ^2 R_ \odot   \times
\frac{1}{{4\pi }}\frac{{\left\langle {\sigma v} \right\rangle
}}{{2m_\chi ^2 }}\frac{{dN}}{{dE}} \times J_{\Delta \Omega }^A
\label{ADMfluxAve}
\end{equation}
\begin{equation}
\phi _{\Delta \Omega }^D (E) = \rho _ \odot  R_ \odot   \times
\frac{1}{{4\pi }}\frac{1}{{m_\chi  \tau _\chi  }}\frac{{dN}}{{dE}}
\times J_{\Delta \Omega }^D. \label{DDMfluxAve}
\end{equation}

\begin{figure}[h]
\vspace*{-.03in} \centering
\includegraphics[width=4.0in,angle=0]{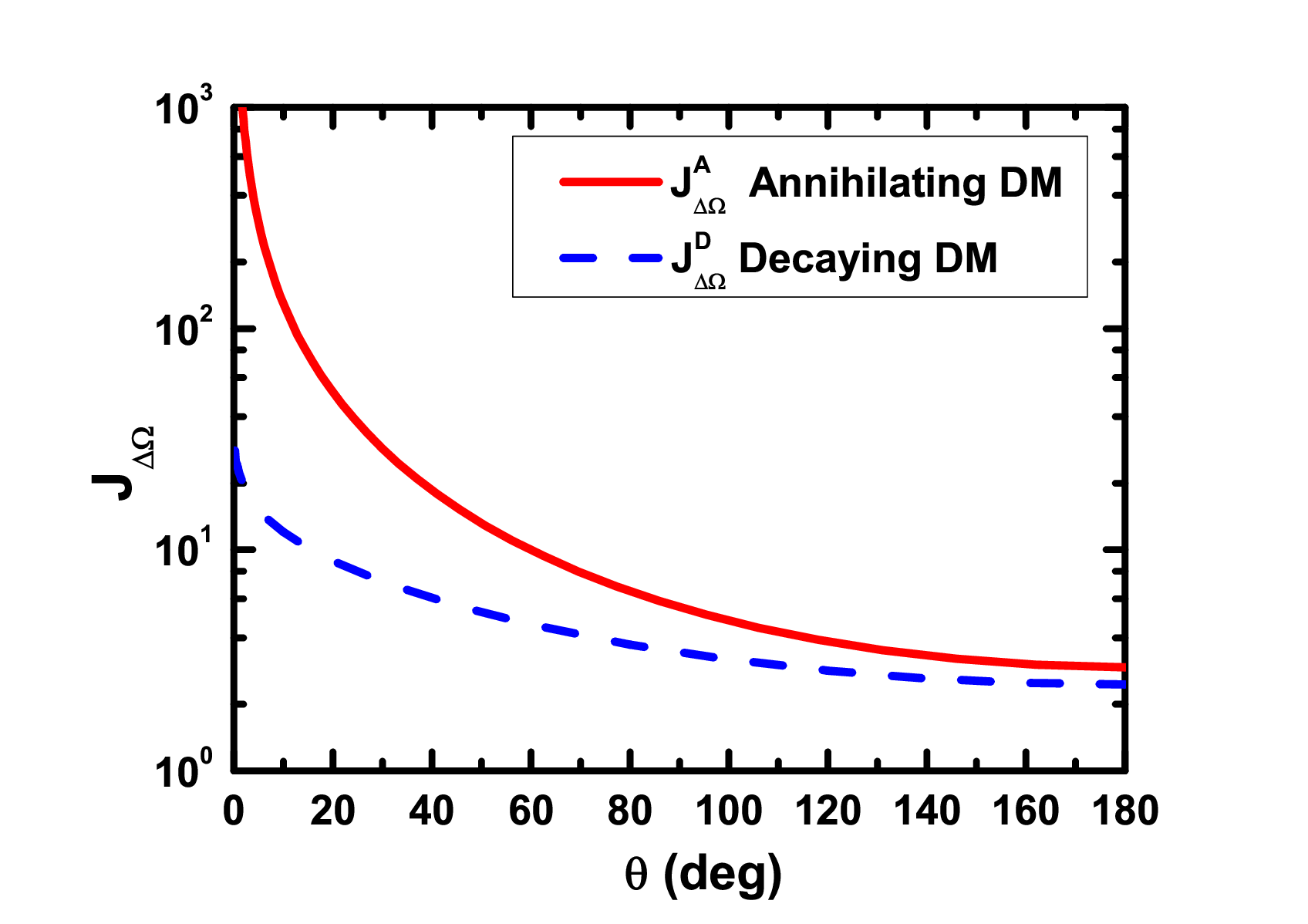}
 \caption{$ J_{\Delta \Omega }$ for
annihilating and decaying DM as the function of half-angle $\theta$
for the cone centered at the direction of the GC. The DM
distribution adopted here is NFW profile.
} \label{Jfactor}
\end{figure}

In order to calculate the neutrino flux, we must know both the
astrophysical $ J_{\Delta \Omega }$ factor and the neutrino spectrum
from charged leptons. In Fig. \ref{Jfactor}, we plot the $ J_{\Delta
\Omega }$ factor for annihilating and decaying DM respectively. We
used NFW \cite{Navarro:1996gj} profile here. From the figure we can
see that the $ J_{\Delta \Omega }$ reach its maximum values, around
1000 and 20 for annihilating and decaying DM, in the direction of GC
as expected. If cusped density like Moore profile is used, one will
get much larger and slightly larger $ J_{\Delta \Omega }$ factors
for annihilating and decaying DM respectively. However, some authors
\cite{Salucci:2002nc} suggested that DM halos around galaxies have
the shallower density than that of predicted by standard cold dark
matter theory.

Besides the GC, subhalos can also be the high energy neutrino
sources. In the direction of the DM subhalos, $ J_{\Delta \Omega }$
factor can be easily as large as 100, or even higher. In order to
keep the line of this paper, we put the analysis on the number
distribution of large $ J_{\Delta \Omega }$ subhalo in the Galaxy in
the Appendix.

In Fig. \ref{PAM-ATIC}, the positron fraction from pure muon and tau
decays, as well as the energy spectrum of positron/electron are
plotted for annihilating and decaying DM. It is a model-independent
discussion of these charged lepton channels, thus we assumed that $
\chi \chi \to l\bar l$ for annihilating DM and $ \chi  \to l\bar l$
for decaying DM respectively. From the figures, we can see that
these channels can account for both the PAMELA and ATIC data with
the proper parameters. In the following calculation of neutrino
flux, we use the same set of parameters to get the neutrino spectrum
$dN/dE$.

\begin{figure}[h]
\vspace*{-.03in} \centering
\includegraphics[width=3.4in,angle=0]{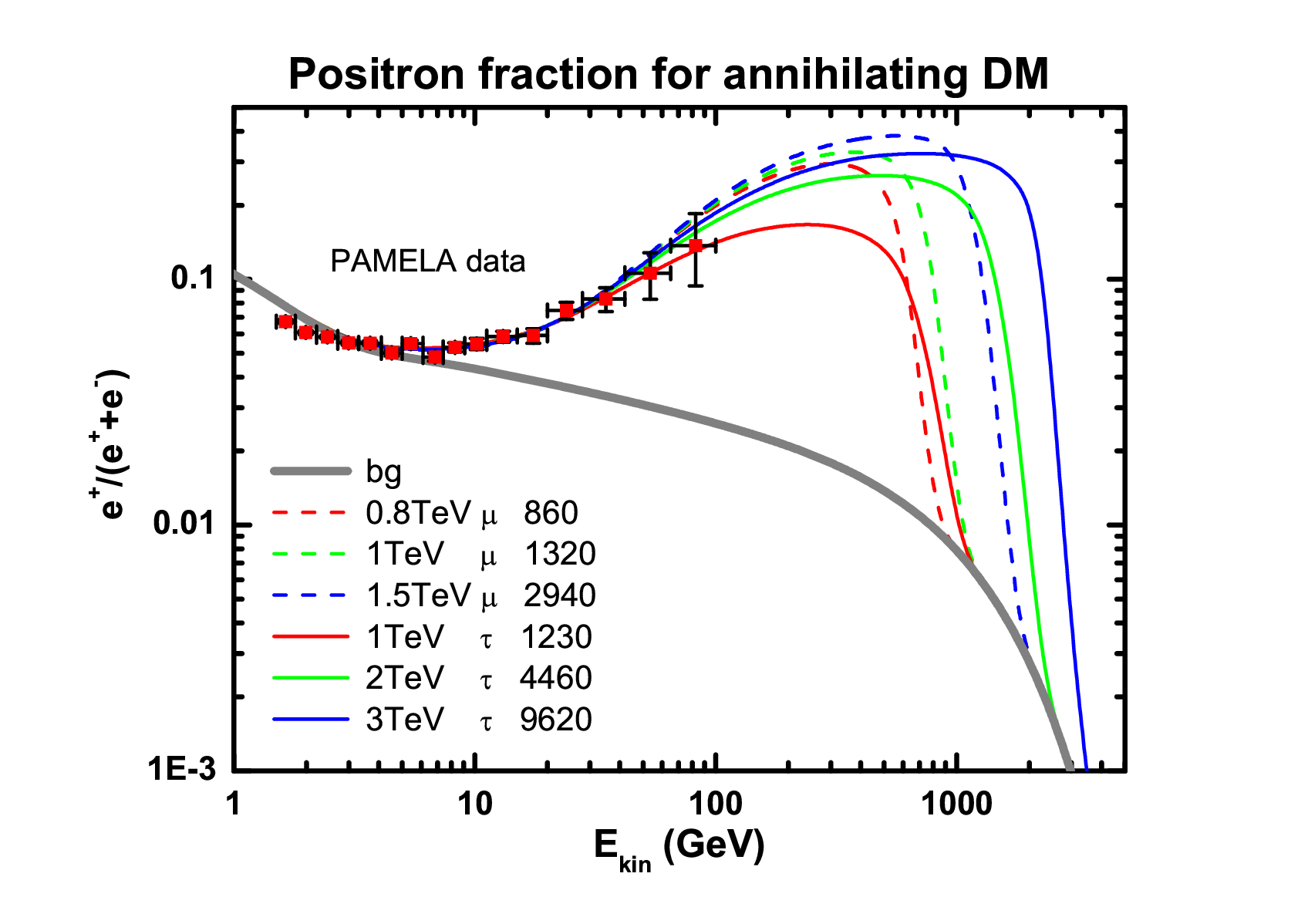}%
\includegraphics[width=3.4in,angle=0]{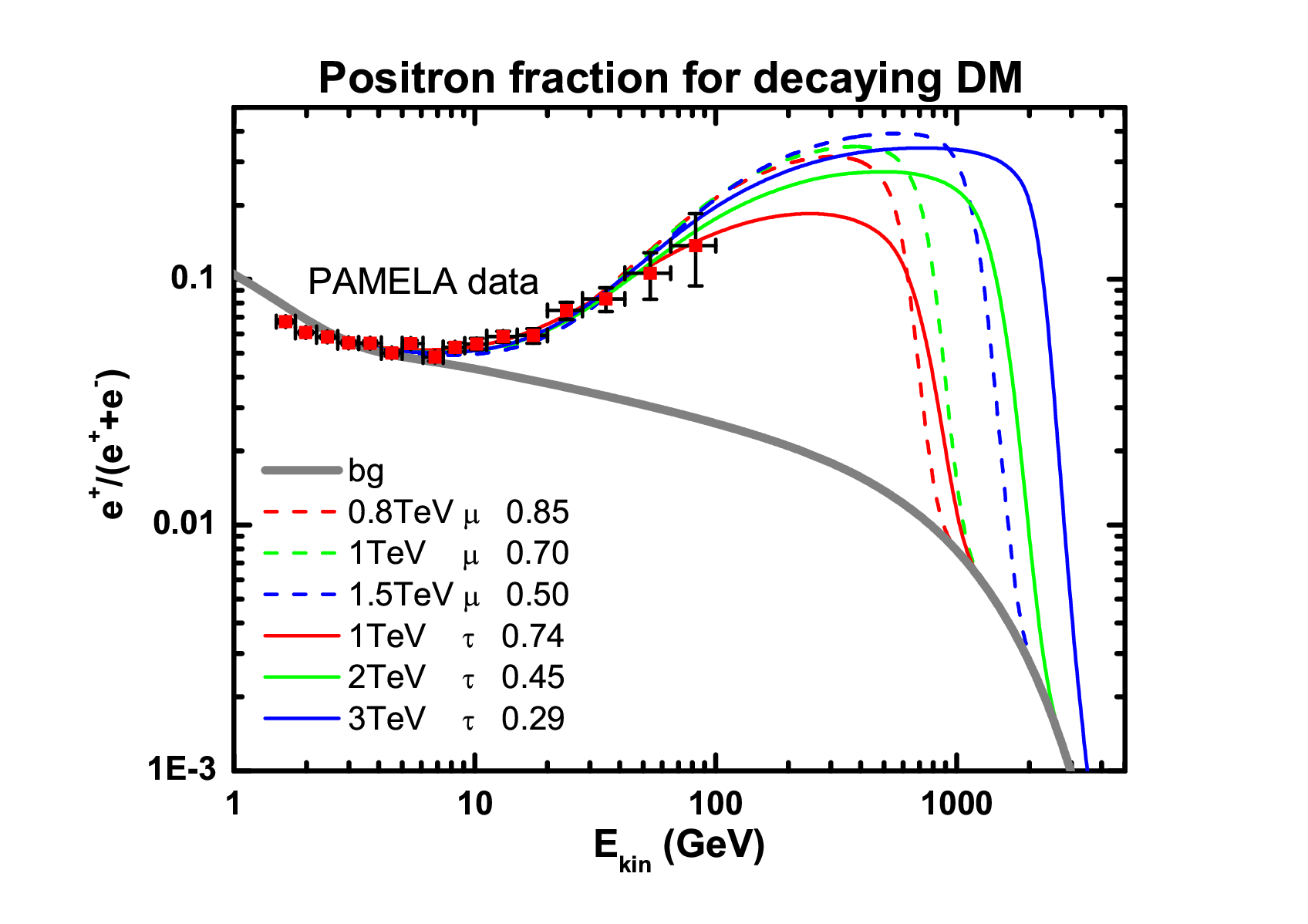}
\\
\includegraphics[width=3.4in,angle=0]{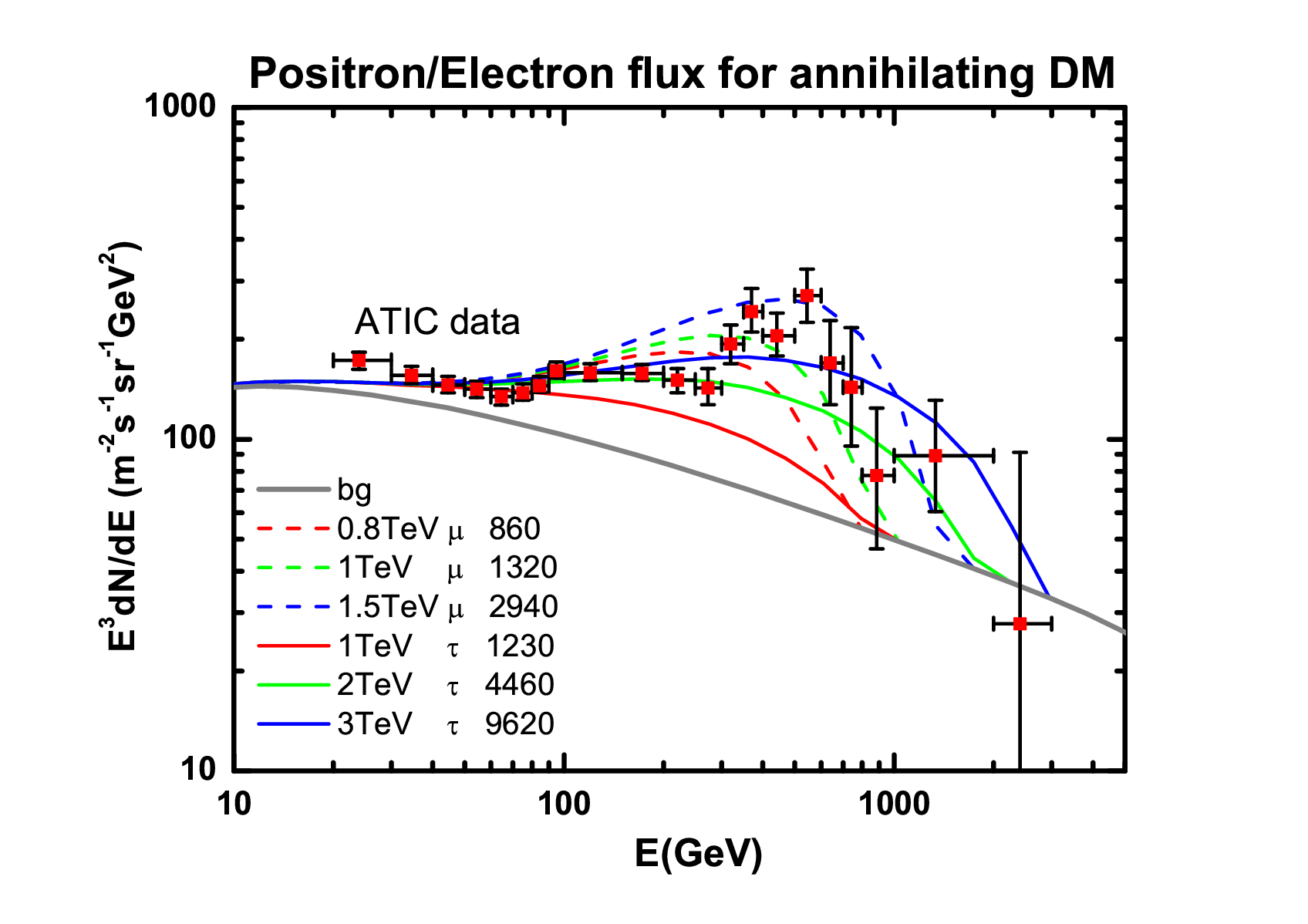}%
\includegraphics[width=3.4in,angle=0]{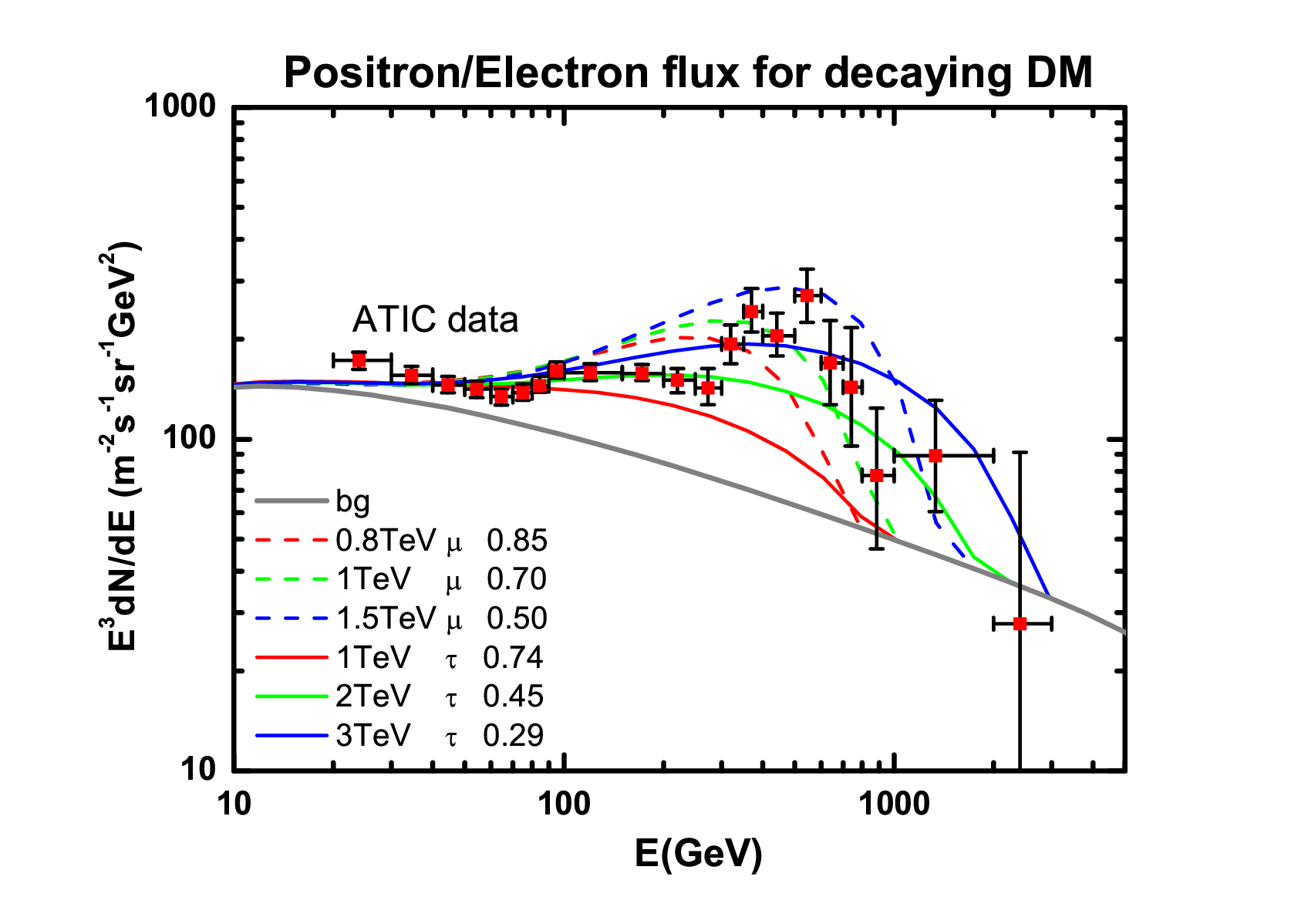}
\vspace*{-.03in} \caption{The positron fraction from pure muon and
tau decays, as well as the positron/electron energy spectrum for
annihilating and decaying DM to account for PAMELA and ATIC data. In
the figures, the labels on the left, e.g. 0.8 TeV,  are the energy
of the muon or tau leptons, while the labels on the right, e.g. 860
or 0.85, are the boost factor or life in unit of $ 10^{26} s$ for
annihilating DM and decaying DM respectively.
} \label{PAM-ATIC}
\end{figure}

\begin{figure}[h]
\vspace*{-.03in} \centering
\includegraphics[width=3.4in,angle=0]{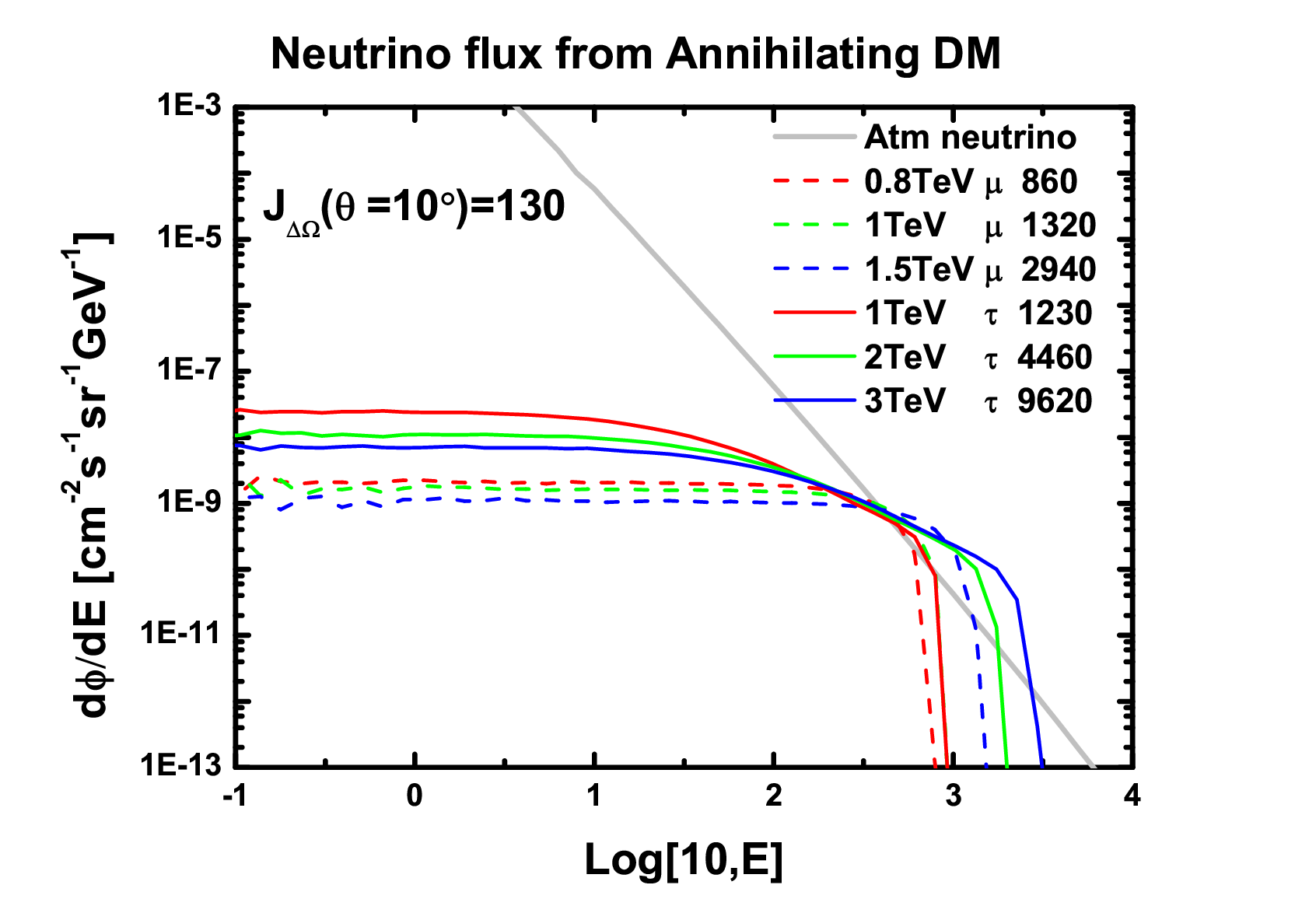}%
\includegraphics[width=3.4in,angle=0]{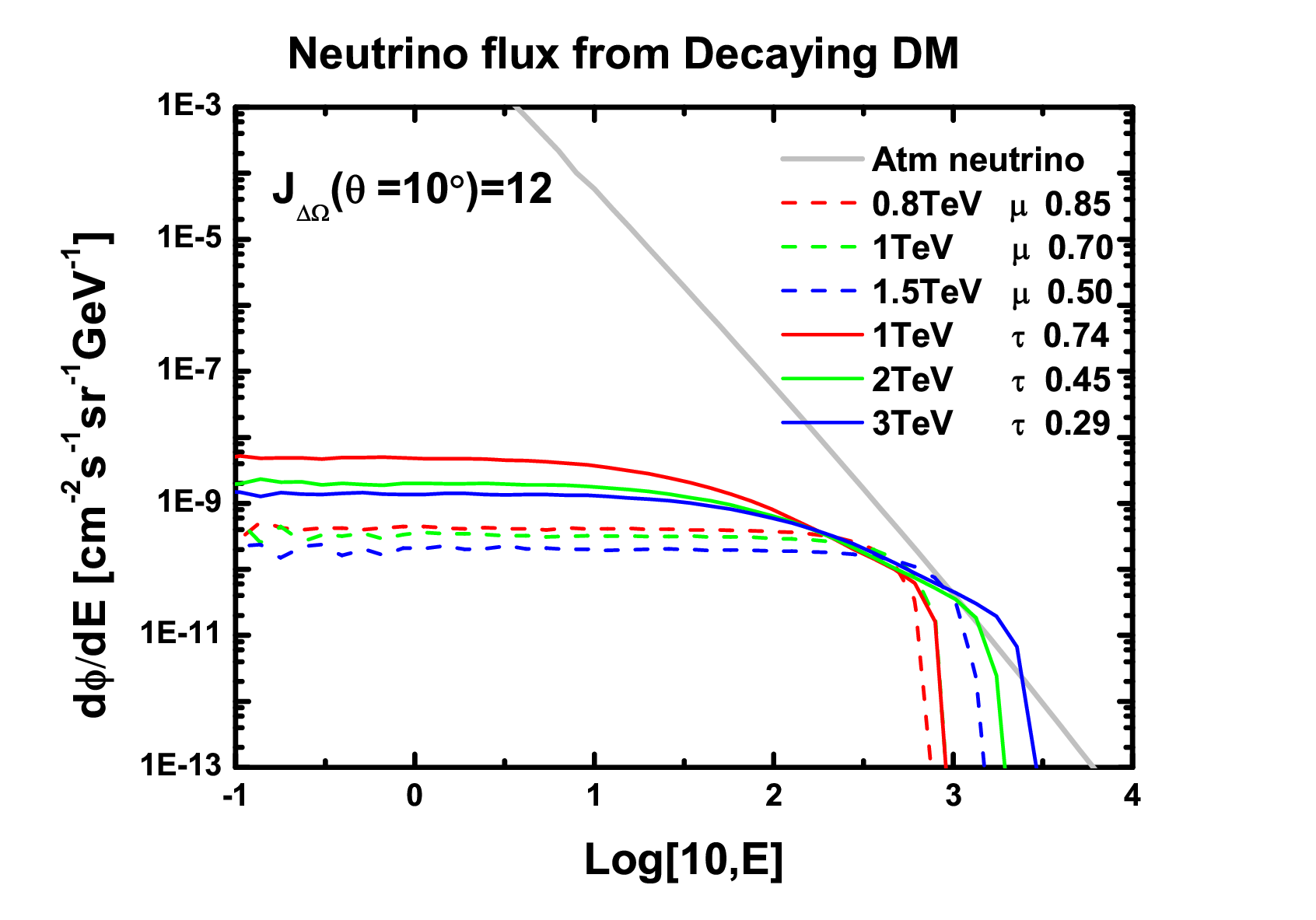}
\\
\includegraphics[width=3.4in,angle=0]{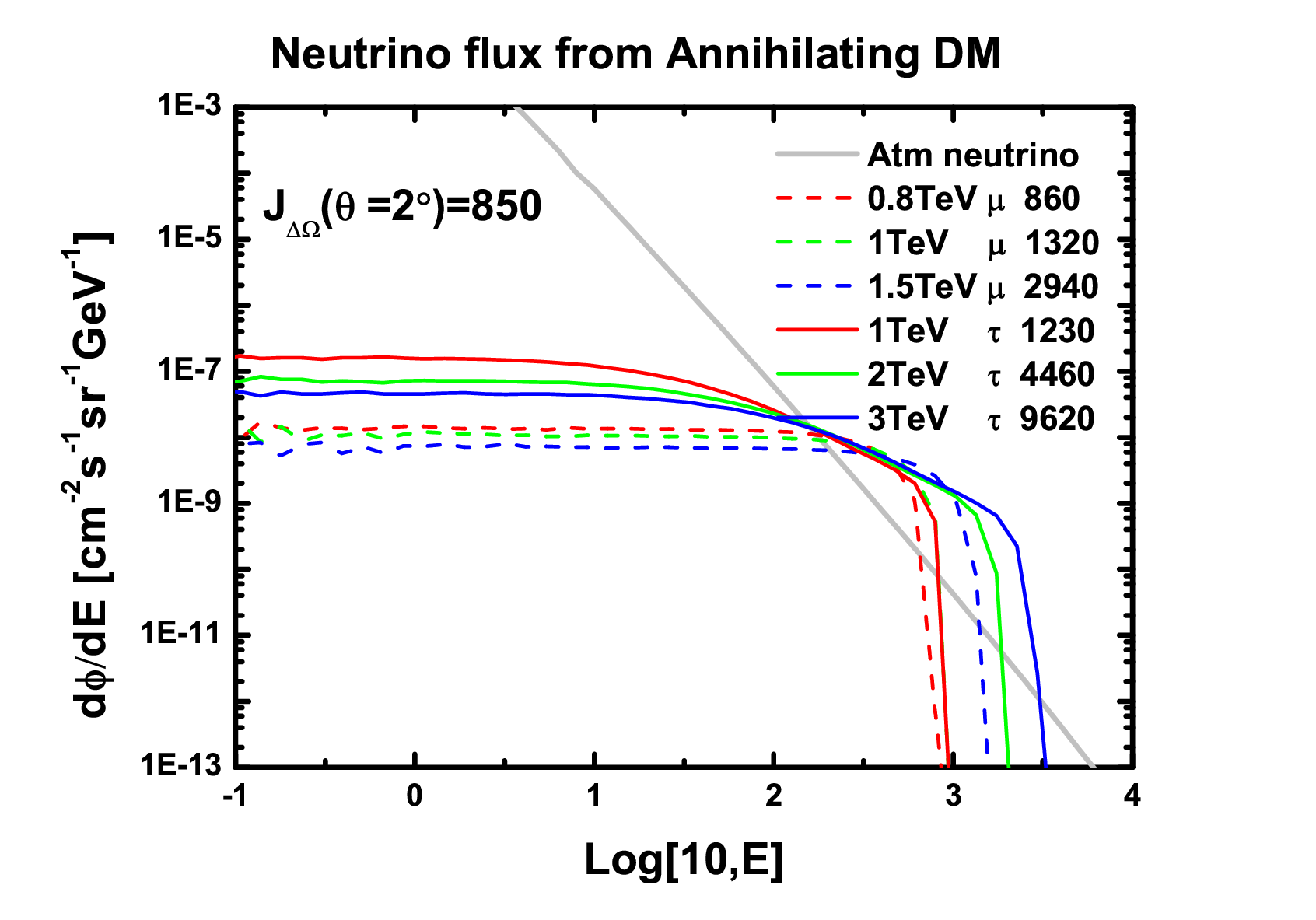}%
\includegraphics[width=3.4in,angle=0]{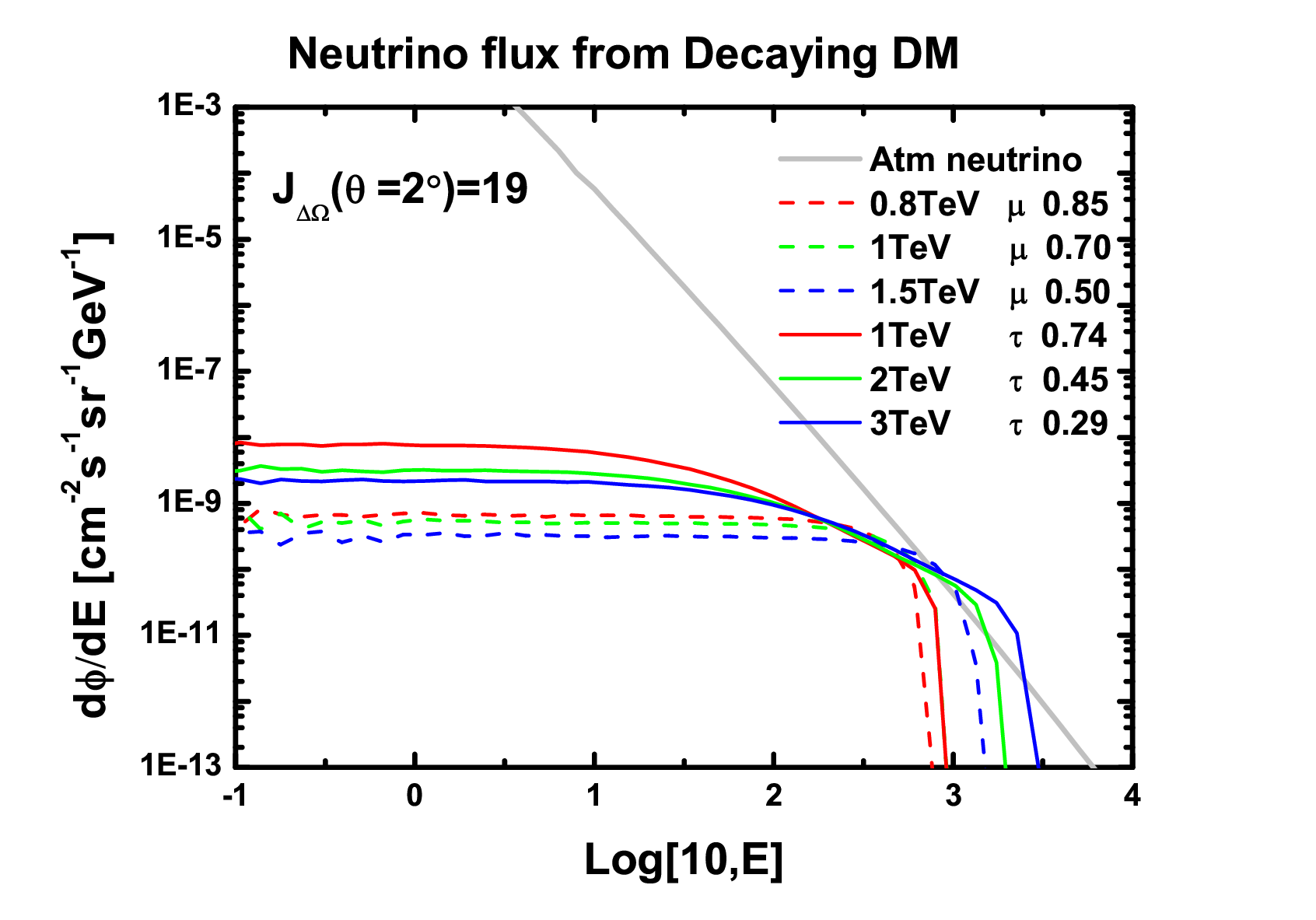}
\vspace*{-.03in} \caption{The muon and anti-muon neutrino flux for
annihilating and decaying DM at the GC. The atmospheric neutrino
flux data is from Ref. \cite{Honda:2006qj}. The x-axis is the energy
of neutrinos and the half angle $ \theta$ is $ 10^ \circ$ and $ 2^
\circ$ respectively. Other labels in the figure are similar with
Fig.~\ref{PAM-ATIC}.
} \label{DMneu}
\end{figure}

Based on the $ J_{\Delta \Omega }$ factor and the neutrino spectrum
from charged leptons, we can calculate the neutrino flux.  In Fig.
\ref{DMneu}, the expected neutrino flux from muon and tau decay for
annihilating and decaying DM from the GC are plotted. We choose the
half angle $ \theta$ to be $ 10^ \circ$ which is the angular
resolution of Super-K for the neutrino energy from 1 GeV to 10 GeV
for a conservative analysis \cite{Ashie:2005ik}. For the neutrino
with higher energy, the Super-K is difficult to measure the energy
of the muons but can count the number. The parameters BF and life of
DM are taken to be the values which can account for the PAMELA and
ATIC data. The figures show that the two DM scenarios have
comparable average neutrino flux with that of the atmospheric
neutrino \cite{Honda:2006qj} for high energy neutrino, because of
the logarithmic decrease of the atmospheric neutrino. For the low
energy region, the neutrino flux are much smaller than that of the
atmospheric neutrino. For some heavy DM, the Super-K can even place
the constraints \cite{Beacom:2006tt,Liu:2008kz} on their
annihilating cross section or life \cite{PalomaresRuiz:2007ry}. For
the smaller cone with $\theta= 2^ \circ$, high energy neutrino from
the annihilating DM have much larger flux than that from the
background.

In the subhalo, the Sommerfeld enhancement for the DM annihilation
can be larger than that in the halo because the relative velocity of
the DM particles are lower. For example the subhalos can have the
velocity dispersion of $O(10) km/s$ \cite{Strigari:2006rd}, which is
smaller than the usual velocity of DM $ v \sim 10^{ - 3} c \sim 300
km/s$. Thus one can expect an extra Sommerfeld enhancement for the
subhalo \cite{ArkaniHamed:2008qn,Lattanzi:2008qa}. The larger BF in
the subhalo will not contradict with the current atmospheric data,
since this effect will be diluted with the larger cone angle where J
factor decreases rapidly. In this paper we do not adopt this extra
enhancement, however we note here that such enhancement can make the
discovery of neutrino from subhalo in annihilating DM scenario even
more promising. The decaying DM have much smaller J factor for
massive subhalo, since it does not benefit from the cusped profile
very much. Thus it is possible to detect neutrino from subhalo in
annihilating DM scenario, but very challenging in decaying DM
scenario.

\section{Muon rate at neutrino telescopes}

The neutrino detector Antares have angular resolution of $ 2^ \circ$
from 100GeV to 1TeV \cite{Stolarczyk:2007ew} and can see GC for $
63\%$ of a year \cite{Aslanides:1999vq}. The Antares has about eight
years of operation time from 2007 to 2015.  Using the neutrino
effective area $ A_v^{eff} (E)$ of Antares \cite{Bruijn:2008zz}, we
calculate the events in the $ 2^ \circ$ cone in the
Table.~\ref{tabAntares} for annihilating DM. The total number of
events is $ N_v  = \int A _v^{eff} (E)\frac{{d\phi _v }}{{dE}}dE$.
From the table, we can see that Antares may discover the neutrino
signal from GC in annihilating DM scenario, but difficult in
decaying DM scenario because the signal gets negligible increase
from smaller cone, which can be seen in the lower panel of Fig.
\ref{DMneu}. The future neutrino detector like KM3NeT can further
explore the annihilating DM scenario.

\begin{table}[htb]
\begin{tabular}{||c|c|c||c|c|c||}
\hline \hline
channel & N   & $\sigma$    & channel   & N   & $\sigma$   \\
\hline
  atm  & 1.5 & $ -$        & atm& 1.5 & $ -$   \\
\hline
  0.8TeV $\mu$ & 7.7  & 6.2 & 1TeV $\tau$& 12.2&  9.9   \\
  \hline
  1TeV $\mu$   & 16.5 & 13.4& 2TeV $\tau$& 21.2& 17.2  \\
  \hline
  1.5TeV $\mu$ & 29.4 & 23.9& 3TeV $\tau$& 23.3& 18.9   \\
\hline \hline
\end{tabular}
\caption{The neutrino event numbers in the energy interval $ 500GeV
- 1TeV$ for eight years of Antares operation from the $ 2^ \circ$
cone in the GC direction. $ \sigma$ is the significance defined as $
S/\sqrt B$.
 \vspace*{-.1in}}\label{tabAntares}
\end{table}

IceCube locates in the south pole and covers the northern sky. It is
excellent to look at the possible large DM subhalo in the Galaxy,
provided that it has good angular resolution like $ 1^ \circ$
\cite{Ahrens:2003ix}, which can greatly suppress the atmospheric
neutrino background. The massive DM subhalo can have high density in
the center, thus it can produce large neutrino flux and can be
identified by the high resolution neutrino detector like IceCube.
Subhalo can easily reach large J factor, say $ J_{\Delta \Omega
}^{Subhalo} (\theta  = 1^ \circ ) \sim 100$ or even larger values
\cite{Yin:2008mv}. In the Appendix, the number distribution of
massive DM subhalos in our Galaxy has been discussed in detail. In
the following estimation, we do not adopt the extreme subhalo model,
instead we take J factor equal to 100.

For the IceCube, we use the same method in Ref.
\cite{Barger:2007xf,Liu:2008kz} to estimate the muon rate in the
neutrino telescopes in annihilating DM scenario. The total muon and
anti-muon rate is expressed as following,
\begin{equation}
\frac{{dN_\mu  }}{{dE_\mu  }} = \int_{E_\mu  }^\infty  {\frac{{d\phi
_{\nu _\mu  } }}{{dE_{\nu _\mu  } }}} \left[\frac{{d\sigma _\nu ^p
(E_{\nu _\mu  } ,E_\mu  )}}{{dE_\mu  }}\rho _p  + \frac{{d\sigma
_\nu ^n (E_{\nu _\mu  } ,E_\mu  )}}{{dE_\mu  }}\rho _n \right]R_\mu
(E_\mu )A_{eff} (E_\mu  )dE_{\nu _\mu  }  + (\nu  \to \bar \nu ),
\label{muonrate}
\end{equation}
where $ d\phi _{\nu _\mu  } /dE_{\nu _\mu  } $ is the muon neutrino
flux arrived at the neutrino detector. $d\sigma _\nu ^p (E_{\nu _\mu
}, E_\mu )/dE_\mu $ and $d\sigma _\nu ^n (E_{\nu _\mu }, E_\mu
)/dE_\mu $ are differential cross sections for the muon production
process $ vp \to lX$ and $ vn \to lX$. The densities of protons and
neutrons near the detector are taken to be $ \rho _p  =
\frac{5}{9}N_A cm^{ - 3}$ and $ \rho _n = \frac{4}{9}N_A cm^{ - 3}$
respectively for IceCube where detector volume is filled with ice. $
N_A$ is the Avagadro's number. The muon energy loss is given as $
\frac{dE}{dx}=-\alpha-\beta E$, where $\alpha$ and $\beta$ are
empirical parameters. The distance that a muon travels in the Earth
before its energy drops below threshold energy $E^{thr}$, is called
muon range which is given as
\begin{equation}
R_{\mu}(E)=\frac{1}{\rho\beta}\ln(\frac{\alpha+\beta E}{\alpha+\beta
E^{thr}}),
\end{equation}
where we take the parameters $ \alpha  = 2.0 \times 10^{ - 6}
TeVcm^2 /g$
 and $
\beta  = 4.2 \times 10^{ - 6} cm^2 /g$
 to be the same
as Ref. \cite{Beacom:2006tt}. The effective area of neutrino
telescope $ A_{eff}$ is the function of muon energy which increases
as the energy goes up. The symbol $ (\nu \to \bar \nu )$ means that
we also count the anti-muons from the anti-muon neutrinos because
the Cherenkov neutrino telescope detects both muons and anti-muons.

\begin{figure}[h]
\vspace*{-.03in} \centering
\includegraphics[width=3.4in,angle=0]{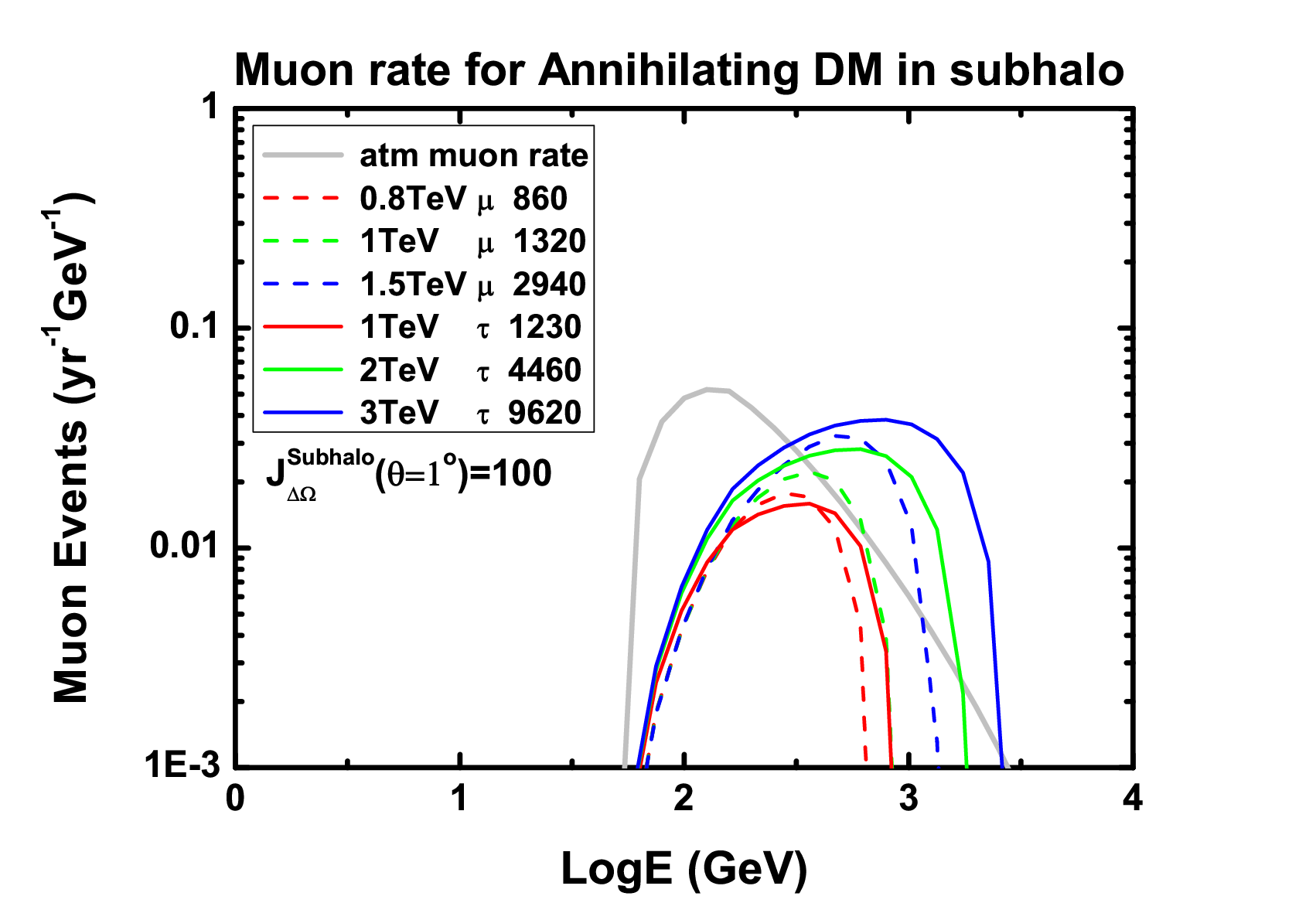}%
\vspace*{-.03in} \caption{The total muon and anti-muon rates for
annihilating DM in the massive subhalo for the IceCube. The half
angle $ \theta$ is taken to be $ 1^ \circ$.
} \label{ADMmuonSubhalo}
\end{figure}

In Fig.~\ref{ADMmuonSubhalo}, we show the muon rate as a function of
muon energy for several parameters which can account for PAMELA and
ATIC data. From the figures we can see that the signal from
annihilating DM can be larger than the atmospheric background for
heavy DM, even without the extra enhancement from lower velocity
dispersion. We calculate the muon number in the energy interval $
500GeV\sim1TeV$ for ten years and give the significance of signal $
\sigma  = S/\sqrt B$ in the Tab.~\ref{tabIce}. Heavy DM can reach
the five sigma significance which is quite encouraging for subhalo
neutrino search in the annihilating DM scenario. On the other hand,
if no neutrino hot spots are identified, it can put constraints on
subhalo model and BF for annihilating DM.

\begin{table}[htb]
\begin{tabular}{||c|c|c||c|c|c||}
\hline \hline
channel    & N   & $\sigma$    & channel   & N   & $\sigma$   \\
\hline
  atm          & 57.6 & $-$    &  atm       & 57.6 & $ -$   \\
\hline
  0.8TeV $\mu$ & 21.7 & 2.9    & 1TeV $\tau$& 41.5&  5.5   \\
  \hline
  1TeV $\mu$   & 55.2 & 7.3    & 2TeV $\tau$& 136.4& 20.0  \\
  \hline
  1.5TeV $\mu$ & 144.9&19.1    & 3TeV $\tau$& 188.6& 24.8   \\
\hline \hline
\end{tabular}
\caption{The total muon and anti-muon numbers in the energy interval
$ 500GeV - 1TeV$ for ten years operation of IceCube for massive
subhalo. $ \sigma$ is the significance defined as $ S/\sqrt B$.
 \vspace*{-.1in}}\label{tabIce}
\end{table}

Finally we would like to discuss the limits from the Super-K
observation. Super-K have searched neutrino signals from DM
annihilation in the direction of the Sun, the Earth and the GC.
Currently, no high energy astrophysical neutrinos have ever been
detected. Super-K collaboration detected the upward going muons from
several $GeV$ to $10 TeV$ in the direction of the center with half
angles from 5 to 30 degrees, and set limit on muon flux
\cite{Desai:2004pq}. For example, Super-K collaboration placed an
upper bound of muon flux in the half angle $ 10^ \circ$ around the
direction to the GC as $\sim 5\times 10^{-15}cm^{-2}s^{-1}$. In our
numerical examples for annihilating DM, the muon flux are $ 4.2
\times 10^{ - 15} cm^{ - 2} s^{ - 1}$, $ 6.5 \times 10^{ - 15} cm^{
- 2} s^{ - 1}$, $ 1.4 \times 10^{ - 14} cm^{ - 2} s^{ - 1}$, $ 5.1
\times 10^{ - 15} cm^{ - 2} s^{ - 1}$, $ 1.7 \times 10^{ - 14} cm^{
- 2} s^{ - 1}$ and $ 3.4 \times 10^{ - 14} cm^{ - 2} s^{ - 1}$ for
0.8TeV $ \mu$, 1TeV $ \mu$, 1.5TeV $ \mu$, 1TeV $ \tau$, 2TeV $
\tau$ and 3TeV $ \tau$ respectively. We should emphasize that in our
calculations we have assumed that DM annihilate into pure muons or
taus. If DM can annihilate to electron/positrons, it is easy to fit
PAMELA and ATIC data with much smaller BF for muon and tau channels
and avoid the violation of Super-K limits. On the other hand, the DM
density distribution is still not very clear, especially in the
central region of Galaxy and subhalos. Different DM profiles give
very different contributions to $J_{\Delta \Omega } (\theta )$,
especially when $ \theta$ is small. For a cusped profile, the $
J_{\Delta \Omega } (\theta )$ for small $ \theta$ will be larger and
can compensate the decrease of BF, which can keep the neutrino
signal large enough and avoid the violation of Super-K limits
simultaneously. We would like to mention that more stringent
constraints on the DM profile may come from the detections of gamma
ray and synchrotron radiation in the GC \cite{photon}.

\section{Conclusions and discussions}

Based on the PAMELA and ATIC observations, there may exist extra
sources of charged leptons, namely electron, muon and tau. In this
paper, we calculated the neutrino flux from muon and tau which arise
from the annihilating/decaying DM. The final muon rate at neutrino
telescopes Antares and IceCube are also estimated.
 Our results show that Antares is promising in
discovering the neutrino signal from GC in annihilating DM scenario,
but challenging in decaying DM scenario. Moreover, massive DM
subhalo is also the promising high energy neutrino source in the
annihilating DM scenario. For ten years operation, IceCube can reach
five sigma significant neutrino signal from subhalos for heavy DM.
Note that in the subhalo the extra Sommerfeld enhancement from lower
velocity dispersion can further improve the neutrino signal.

In this paper we focus on the muon neutrino detection. Actually the
tau neutrino is deserved further investigation because the
atmospheric neutrino contains less tau neutrino than muon neutrino.
If one only counts the tau numbers in the neutrino detector, it can
further suppress the atmospheric background \cite{Covi:2008jy} and
this can be done in future kilometer neutrino detector like KM3NeT.
Finally, it is worthy to mention that pulsars are difficult to be
detected by upcoming neutrino detector \cite{Bhadra:2008cg}, thus
one may distinguish it from the DM scenarios.

\acknowledgements

J. Liu and P. F. Yin thank N. Weiner for useful discussions and
encouragement. We thank X. J. Bi for useful discussions. This work
was supported in part by the Natural Sciences Foundation of China
(No. 10775001 and 10635030).

\emph{Note added:} During the completion of this work one similar
analysis appeared \cite{Hisano:2008ah}. As a cross check, we have
calculated the muon flux in the 10 degree half-angle cone from the
GC for various channels, which agree with their results.

\appendix

\section{Massive subhalo with large $
J_{\Delta \Omega }$ }

In order to determine the subhalo DM contributions to neutrino flux,
generally speaking, we need to know the DM profile in the subhalo
and the number density of subhalos in the Galaxy. Currently N-body
simulations provide useful information about the DM distribution in
the Galaxy. In this Appendix, we utilize the models based on N-body
simulations to investigate the number distribution of the massive DM
subhalo with large $J_{\Delta\Omega}$, which is defined in Eqs.
(\ref{ja}) and (\ref{jadomega}).

\subsection{DM subhalo profile}

Based on  N-body simulations, the DM distribution can usually be
parameterized as,
\begin{equation}
\rho(r)=\frac{\rho_s}{(r/r_s)^{\gamma}[1+(r/r_s)^{\alpha}]^{(\beta-\gamma)/\alpha}},
\label{nfwmoore}
\end{equation}
where $\rho_s$ and $r_s$ are the scale density and scale radius
parameters respectively. The parameters $(\alpha,\beta,\gamma)$ are
$(1,3,1)$ and $(1.5,3,1.5)$ for NFW \cite{Navarro:1996gj} and Moore
\cite{Moore:1999gc} profiles respectively.

The two free parameters $\rho_s$ and $r_s$ for a subhalo can be
determined once we know the subhalo mass $M_v$ and the concentration
parameter $c_v$ which depends on the specific subhalo model. The
density scale $\rho_s$ can be determined by using the mass relation
\begin{equation} \int\rho_s(r) {\rm d}V=M_v. \end{equation}
In the following we will concentrate on how to determine $r_s$.

If $M_v$ is known, we can calculate the virial radius $r_v$ of the
subhalo which is often approximated as the radius within which the
average density is greater, by a specific factor $\Delta=200$, than
the critical density of the Universe $\rho_c=139$ M$_{\odot}$
kpc$^{-3}$ (M$_{\odot}$ is mass of the Sun). Thus $r_v$ can be
expressed as \begin{equation}
 r_v = \left( {\frac{{M_v }}{{(4\pi
/3)\Delta \rho _c }}} \right)^{1/3}. \end{equation} $r_v$ describes
the radius within which the subhalo can hold together by its own
gravity, but it does not describe how the DM mass are distributed
inside the distance $r_v$. On the other hand, the scale radius $r_s$
describes the DM mass distribution in the subhalo that most of mass
are within this radius. The concentration models assume that the
$r_s$ is proportional to the $r_v$ and determine ratio of them.
Large ratio of $ r_v /r_s$ represents the mass of subhalo are highly
concentrated in a small radius in the center, which results in the
cusped profile.

For a concentration model the ratio of $ r_v /r_s$ is closed related
to concentration parameter $c_v$, which is defined as $
c_v=\frac{r_v}{r_{-2}}$. Here $r_{-2}$ is another radius parameter
which is defined as $ \frac{d}{dr}(r^2 \rho)|_{r=r_{-2}}=0$.
$r_{-2}$ represents the radius within which the mass of the subhalo
is concentrated. To illustrate this point, we can easily check that
$r_{-2}$ is in fact the inflexion point of the mass, i.e. $ \left.
{\frac{{d^2 M(r)}}{{dr^2 }}} \right|_{r = r_{ - 2} }  = 0$ with $
M(r) = \int_0^{r} {\rho (r^\prime)} \cdot 4\pi r^{\prime 2}
dr^\prime$. The relation between $r_{-2}$ and  $r_s$ can be
calculated by the definition of $r_{-2}$ and Eq. (\ref{nfwmoore}).
The $r_s$ for NFW and Moore profiles are $r_s^{nfw}=r_{-2}$ and
$r_s^{Moore}=r_{-2}/0.63$ respectively. Thus the $r_s$ can be
determined by $M_v$, i.e.
\begin{equation} r_s^{nfw}=\frac{r_v(M_v)}{c_v(M_v)}, \ \ \
r_s^{moore}=\frac{r_v(M_v)}{0.63\,c_v(M_v)}. \end{equation} Based on
the above description, we can see that once the $c_v-M_v$ relation
is given by the concentration model, the DM profile of subhalo is
determined.

We use the same method as Ref. \cite{Yin:2008mv} which adopts two
concentration models, which are ENS01 \cite{eke01}  and B01
\cite{bullock01}. In the Ref. \cite{lavalle08}, the $c_v$ is fitted
in a polynomial form as
\begin{equation}
\ln(c_v)=\sum_{i=0}^4C_i\times\left[\ln\frac{M_v}{M_{\odot}}\right]^i,
\label{lncv}
\end{equation}
where $C_i=\{3.14,-0.018,-4.06\times10^{-4},0,0\}$ and
$\{4.34,-0.0384,-3.91\times10^{-4},-2.2\times10^{-6},
-5.5\times10^{-7}\}$ for ENS01 and B01 model respectively.

\begin{figure}[!htb]
\begin{center}
\includegraphics{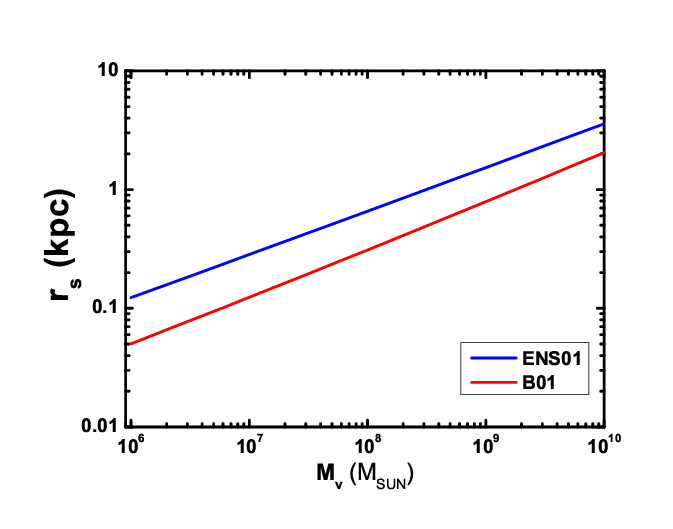}
\caption{The scale radius $r_s$ as a function of subhalo mass
$M_{v}$.} \label{rsNFW}
\end{center}
\end{figure}

In Fig. \ref{rsNFW}, the scale radius $r_s$ as a function of subhalo
mass $M_{v}$ is plotted to show how large the subhalos are. The
subhalo radius is smaller in B01 model than that in ENS01 model
because the $c_v$ is larger in B01 model. This plot assumes that
subhalos have NFW profile. For the Moore profile $r_s^{moore}$ is
equal to $r_s^{nfw}/0.63$.

\subsection{Subhalo number distribution}

 The number of subhalos has been revealed by
many N-body simulations \cite{sim}. The number density of subhalos
have a power-law relation with subhalo mass and an uniform
distribution in the solid angle as,
\begin{equation}
\frac{{\rm d}N}{{\rm d}M_{v}\cdot4\pi r^2{\rm d}r}=
N_0\left(\frac{M_{v}}{M_{host}}\right)^{-\alpha}
\frac{1}{1+\left(\frac{r}{r_H}\right)^2}, \label{number}
\end{equation}
where $r$ is the distance from the GC to subhalo, $M_{host}$ is the
mass of host halo (the virial mass of the Galaxy of
$M_{host}=10^{12}M_{\odot}$). $r_H$ is the core radius which is
usually a fraction of the virial radius of the host halo (The
relation $r_H\approx0.14 r_v$ is adopted here
\cite{Diemand:2004kx}). The slope $\alpha$ varies from 1.7 to 2.1,
and an intermediate $\alpha$ is adopted, i.e. $\alpha = 1.9$
\cite{Madau:2008fr}. $N_0$ is the normalization factor which is
determined by setting the number of subhalos with mass larger than
$10^8 M_ \odot$ to be 100 \cite{lavalle08}. Thus if $M_{v}$ and the
distance $r$ of the subhalo are known, we can calculate the number
density of the subhalos.

\subsection{Number of subhalos with $
J_{\Delta \Omega (1^ \circ  )}  > 100$ }

In this subsection, we will discuss the number of massive subhalo
with $ J_{\Delta \Omega (1^ \circ  )} > 100$ in our Galaxy. From the
above discussions, once the concentration model (equivalently
saying, the function $ c_v (M_v )$ ) is given, the
$J_{\Delta\Omega}$ is determined by $M_{v}$ and the subhalo position
in the Galaxy. Generally speaking, the heavier subhalo will have
larger $J_{\Delta\Omega}$. At the same time, the distance between
the Earth and the subhalo is closer, the $J_{\Delta\Omega}$ will be
larger.

A monte carlo method is adopted to calculate the number distribution
of DM subhalos with mass larger than $10^6 M_{\odot}$, according to
the distribution function Eq. (\ref{number}). Besides $M_{v}$ and
subhalo distance $r$ from GC, we still need the azimuthal angles
$\theta$ and $ \phi$, which are uniformly distributed in the solid
angle to specify the position of the subhalo.
Once we know the above four parameters, the $ J_{\Delta \Omega }$
can be calculated for any specific concentration model. The
concentration models of ENS01 and B01 are adopted in the
calculations, while the DM profiles of subhalos used here are the
NFW and Moore ones. The total number of subhalos $ N( > 10^6 M_
\odot)$ are calculated to be about 6400 in our Galaxy for NFW-ENS01,
Moore-ENS01, NFW-B01 and Moore-B01 subhalo distributions.

\begin{figure}[!htb]
\begin{center}
\includegraphics{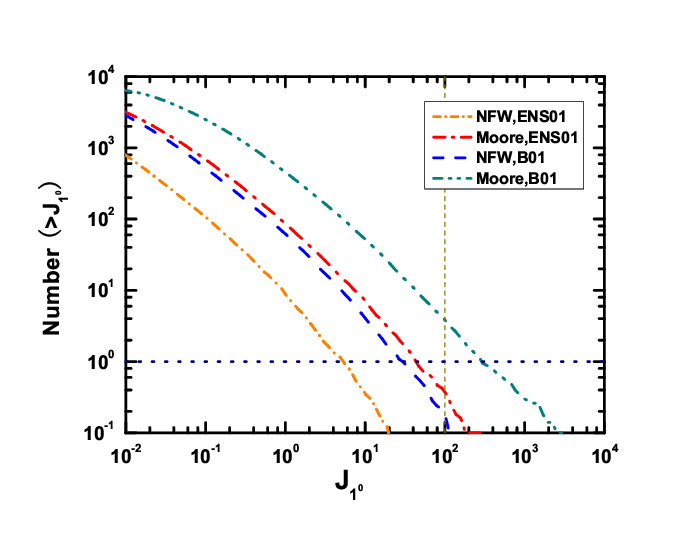}
\caption{The cumulative number of sub-halos with astrophysical
factor $> J_{\Delta \Omega (1^ \circ  )}$ as a function of $
J_{\Delta \Omega (1^ \circ  )}$. The dashed horizon line corresponds
to the case that only one subhalo will be observed.
}\label{jpsisubeps}
\end{center}
\end{figure}

In Fig. \ref{jpsisubeps}, we give the cumulative number of subhalos
with $ J_{\Delta \Omega (1^ \circ  )}$ lager than a specified value.
From the figure we can see the number of subhalos with $ J_{\Delta
\Omega (1^ \circ  )}  > 100$ depends on the concentration model and
subhalo DM profile. Moore-B01 model provides the largest number of
subhalos with high luminosity. There are several subhalos which are
$ J_{\Delta \Omega (1^ \circ  )}  > 100$ in our galaxy in this
model. For the Moore-ENS01 and NFW-B01 model, the subhalo number
with $ J_{\Delta \Omega (1^ \circ )}  > 100$ in 6400 subhalos is
about $ 0.6$ and $ 0.1$ respectively and even smaller for the
NFW-ENS01 model.

\begin{figure}[h]
\vspace*{-.03in} \centering
\includegraphics[width=3.4in,angle=0]{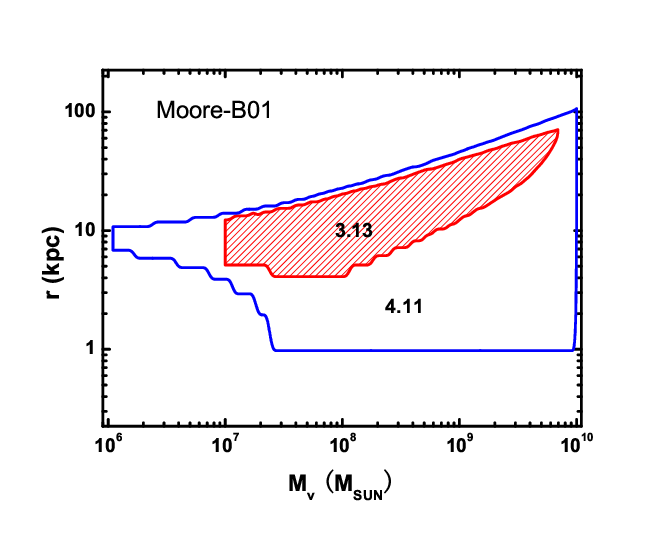}%
\includegraphics[width=3.4in,angle=0]{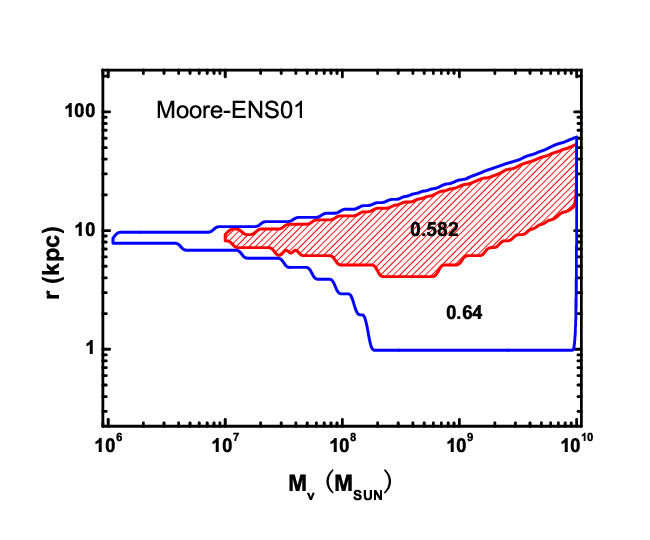}
\\
\includegraphics[width=3.4in,angle=0]{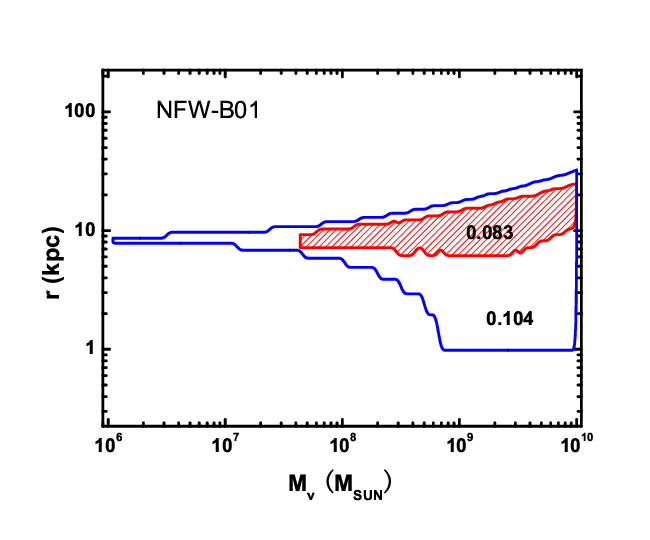}%
\includegraphics[width=3.4in,angle=0]{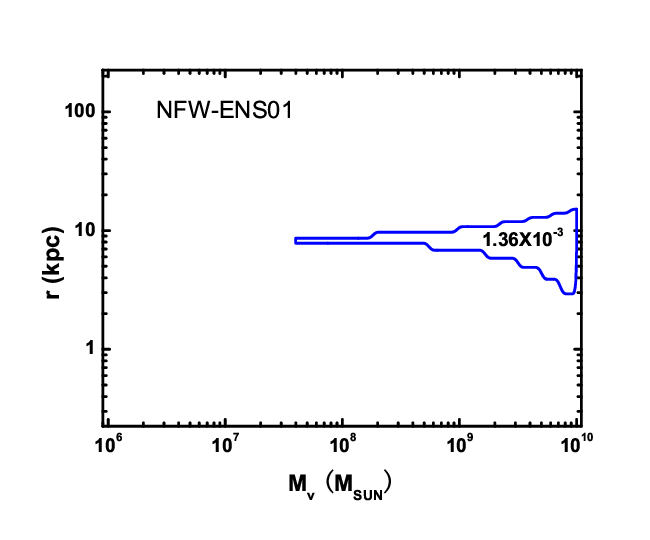}
\vspace*{-.03in} \caption{The number distribution of  subhalos with
$ J_{\Delta \Omega (1^ \circ )}  > 100$. The $r$ is the distance
from the GC, the $M_v$ is the subhalo mass.
} \label{prob}
\end{figure}

In order to show clearly the mass and location of subhalo with $
J_{\Delta \Omega (1^ \circ )}
> 100$, we depict in Fig. \ref{prob} the number distribution in the $r-M_v$ map.
Just for illustration purpose, the quantity inside the blue contour
line shows the total number of these massive subhalos in the whole
$M_v-r$ map. Outside the blue contour line, the number of such
subhalos is zero. Moreover the quantity in the red contour line
shows the subhalo number in the shadowed region. The figure shows
that the subhalos tend to have moderate large mass. It is quite easy
to understand. On one hand, subhalos with larger mass will induce
higher $ J_{\Delta \Omega }$. On the other hand, the number of
higher mass subhalos is less. Thus the competition between these two
effects determines that subhalos with moderate mass tend to provide
the largest flux. The figure also indicates that subhalos tend to
locate around $ 8.5$ kpc from the GC, which is the distance between
the GC and the Earth.

\subsection{Summary}

In this Appendix, we showed that the subhalos with $ J_{\Delta
\Omega (1^ \circ  )} > 100$ are very likely to have intermediate
mass from $ 10^8 M_ \odot$ to $ 10^{10} M_ \odot$, and locate within
several kpc from the Earth. The number of these subhalos in our
Galaxy is about $ O(1)$ which depends on concentration model and
subhalo DM profile. It should be noted that the ENS01 and B01 models
are for distinct halos in the Universe. In our Galaxy, the subhalos
are in the host halo (Milky Way halo) which is denser than the
Universe background. We can expect that the subhalos in the Milky
Way halo should be more concentrated than the distinct halos. The
simulation in Ref. \cite{bullock01} indeed shows the subhalos in a
host halo have larger concentration parameter $ c_v$ than the
distinct ones with the same mass. Thus we can expect larger number
of high luminosity subhalos in our Galaxy comparing with the numbers
above.

From Fig. \ref{rsNFW} and Fig. \ref{prob}, we can see that the whole
subhalo may be difficult to be covered by the 1 degree half-cone,
which means the subhalo can not be treated as `point source'.
Nevertheless, we are interested in the bright core of subhalo which
can still be processed as point source. The IceCube may discover
these hot spots through survey of the sky map by 1 degree half-cone.


\begin{thebibliography}{99}

\bibitem{Adriani:2008zr}
  O.~Adriani {\it et al.},
  arXiv:0810.4995 [astro-ph];
  arXiv:0810.4994 [astro-ph].

\bibitem{Chang:2008zz}
  J.~Chang {\it et al.},
  Nature {\bf 456}, 362 (2008).

 \bibitem{pulsar}
  D.~Hooper, P.~Blasi and P.~D.~Serpico,
  arXiv:0810.1527 [astro-ph];
  H.~Yuksel, M.~D.~Kistler and T.~Stanev,
  arXiv:0810.2784 [astro-ph].

\bibitem{annidm}
  L.~Bergstrom, T.~Bringmann and J.~Edsjo,
  arXiv:0808.3725 [astro-ph];
  V.~Barger, W.~Y.~Keung, D.~Marfatia and G.~Shaughnessy,
  arXiv:0809.0162 [hep-ph];
  I.~Cholis, L.~Goodenough, D.~Hooper, M.~Simet and N.~Weiner,
  arXiv:0809.1683 [hep-ph].
  M.~Pospelov and A.~Ritz,
  arXiv:0810.1502 [hep-ph];
  M.~Fairbairn and J.~Zupan,
  arXiv:0810.4147 [hep-ph].
  I.~Cholis, D.~P.~Finkbeiner, L.~Goodenough and N.~Weiner,
  arXiv:0810.5344 [astro-ph];
  Y.~Nomura and J.~Thaler,
  arXiv:0810.5397 [hep-ph].
  D.~Feldman, Z.~Liu and P.~Nath,
  arXiv:0810.5762 [hep-ph];
  P.~J.~Fox and E.~Poppitz,
  arXiv:0811.0399 [hep-ph].

\bibitem{Cirelli:2008jk}
  M.~Cirelli and A.~Strumia,
  arXiv:0808.3867 [astro-ph];
  M.~Cirelli, M.~Kadastik, M.~Raidal and A.~Strumia,
  arXiv:0809.2409 [hep-ph].

\bibitem{ArkaniHamed:2008qn}
  N.~Arkani-Hamed, D.~P.~Finkbeiner, T.~Slatyer and N.~Weiner,
  arXiv:0810.0713 [hep-ph].

\bibitem{decaydm}
  P.~f.~Yin, Q.~Yuan, J.~Liu, J.~Zhang, X.~j.~Bi and S.~h.~Zhu,
  arXiv:0811.0176 [hep-ph];
  K.~Ishiwata, S.~Matsumoto and T.~Moroi,
  arXiv:0811.0250 [hep-ph];
  C.~R.~Chen, F.~Takahashi and T.~T.~Yanagida,
  arXiv:0811.0477 [hep-ph];
  K.~Hamaguchi, E.~Nakamura, S.~Shirai and T.~T.~Yanagida,
  arXiv:0811.0737 [hep-ph];
  A.~Ibarra and D.~Tran,
  arXiv:0811.1555 [hep-ph];
  C.~R.~Chen, F.~Takahashi and T.~T.~Yanagida,
  arXiv:0811.3357 [astro-ph].

\bibitem{sommerfeld}
  H.~Baer, K.~m.~Cheung and J.~F.~Gunion,
  Phys.\ Rev.\  D {\bf 59}, 075002 (1999)
  [arXiv:hep-ph/9806361];
  J.~Hisano, S.~Matsumoto, M.~M.~Nojiri and O.~Saito,
  Phys.\ Rev.\  D {\bf 71}, 063528 (2005)
  [arXiv:hep-ph/0412403];
  J.~Hisano, S.~Matsumoto, M.~Nagai, O.~Saito and M.~Senami,
  Phys.\ Lett.\  B {\bf 646}, 34 (2007)
  [arXiv:hep-ph/0610249].
  M.~Cirelli, A.~Strumia and M.~Tamburini,
  Nucl.\ Phys.\  B {\bf 787}, 152 (2007)
  [arXiv:0706.4071 [hep-ph]];
  J.~March-Russell, S.~M.~West, D.~Cumberbatch and D.~Hooper,
  JHEP {\bf 0807}, 058 (2008)
  [arXiv:0801.3440 [hep-ph]];
  J.~D.~March-Russell and S.~M.~West,
  arXiv:0812.0559 [astro-ph].

\bibitem{Lattanzi:2008qa}
  M.~Lattanzi and J.~I.~Silk,
  arXiv:0812.0360 [astro-ph].

\bibitem{itoh}
  N. Itoh, H. Hayashi, A. Nishikawa and Y. Kohyama,
  ApJS 102: 411-424 (1996).

\bibitem{Beacom:2006tt}
  J.~F.~Beacom, N.~F.~Bell and G.~D.~Mack,
  Phys.\ Rev.\ Lett.\  {\bf 99}, 231301 (2007)
  [arXiv:astro-ph/0608090];
  H.~Yuksel, S.~Horiuchi, J.~F.~Beacom and S.~Ando,
  Phys.\ Rev.\  D {\bf 76}, 123506 (2007)
  [arXiv:0707.0196 [astro-ph]].

\bibitem{Barger:2007xf}
  V.~Barger, W.~Y.~Keung, G.~Shaughnessy and A.~Tregre,
  Phys.\ Rev.\  D {\bf 76}, 095008 (2007)
  [arXiv:0708.1325 [hep-ph]];
  V.~D.~Barger, W.~Y.~Keung and G.~Shaughnessy,
  Phys.\ Lett.\  B {\bf 664}, 190 (2008)
  [arXiv:0709.3301 [astro-ph]].

\bibitem{PalomaresRuiz:2007ry}
  S.~Palomares-Ruiz,
  Phys.\ Lett.\  B {\bf 665}, 50 (2008)
  [arXiv:0712.1937 [astro-ph]].

\bibitem{Liu:2008kz}
  J.~Liu, P.~f.~Yin and S.~h.~Zhu,
  Phys.\ Rev.\  D {\bf 77}, 115014 (2008)
  [arXiv:0803.2164 [hep-ph]].

\bibitem{Yin:2008mv}
  P.~f.~Yin, J.~Liu, Q.~Yuan, X.~j.~Bi and S.~h.~Zhu,
  Phys.\ Rev.\  D {\bf 78}, 065027 (2008)
  [arXiv:0806.3689 [astro-ph]].

\bibitem{Covi:2008jy}
  L.~Covi, M.~Grefe, A.~Ibarra and D.~Tran,
  arXiv:0809.5030 [hep-ph].

\bibitem{Desai:2004pq}
  S.~Desai {\it et al.}  [Super-Kamiokande Collaboration],
  Phys.\ Rev.\  D {\bf 70}, 083523 (2004)
  [Erratum-ibid.\  D {\bf 70}, 109901 (2004)]
  [arXiv:hep-ex/0404025].

\bibitem{Ashie:2005ik}
  Y.~Ashie {\it et al.}  [Super-Kamiokande Collaboration],
  Phys.\ Rev.\  D {\bf 71}, 112005 (2005)
  [arXiv:hep-ex/0501064].

\bibitem{Ackermann:2004aga}
  M.~Ackermann {\it et al.}  [The AMANDA Collaboration],
  Phys.\ Rev.\  D {\bf 71}, 077102 (2005)
  [arXiv:astro-ph/0412347];
  M.~Ackermann {\it et al.}  [IceCube Collaboration],
  Astrophys.\ J.\  {\bf 675}, 1014 (2008)
  [arXiv:0711.3022 [astro-ph]].


\bibitem{Kistler:2006hp}
  M.~D.~Kistler and J.~F.~Beacom,
  Phys.\ Rev.\  D {\bf 74}, 063007 (2006)
  [arXiv:astro-ph/0607082];
  J.~F.~Beacom and M.~D.~Kistler,
  Phys.\ Rev.\  D {\bf 75}, 083001 (2007)
  [arXiv:astro-ph/0701751].



\bibitem{Aslanides:1999vq}
  E.~Aslanides {\it et al.}  [ANTARES Collaboration],
  arXiv:astro-ph/9907432.

\bibitem{Ahrens:2003ix}
  J.~Ahrens {\it et al.}  [IceCube Collaboration],
  Astropart.\ Phys.\  {\bf 20}, 507 (2004)
  [arXiv:astro-ph/0305196].

\bibitem{Carr:2007zc}
  J.~Carr {\it et al.}  [KM3NeT Collaboration],
  arXiv:0711.2145 [astro-ph].

\bibitem{Navarro:1996gj}
  J.~F.~Navarro, C.~S.~Frenk and S.~D.~M.~White,
  Astrophys.\ J.\  {\bf 490}, 493 (1997)
  [arXiv:astro-ph/9611107].

\bibitem{Salucci:2002nc}
  P.~Salucci, F.~Walter and A.~Borriello,
  arXiv:astro-ph/0206304.

\bibitem{Honda:2006qj}
  M.~Honda, T.~Kajita, K.~Kasahara, S.~Midorikawa and T.~Sanuki,
  Phys.\ Rev.\  D {\bf 75}, 043006 (2007)
  [arXiv:astro-ph/0611418].

\bibitem{Stolarczyk:2007ew}
  T.~Stolarczyk  [ANTARES Collaboration],
  Nucl.\ Phys.\ Proc.\ Suppl.\  {\bf 165}, 188 (2007).

\bibitem{Strigari:2006rd}
  L.~E.~Strigari, S.~M.~Koushiappas, J.~S.~Bullock and M.~Kaplinghat,
  Phys.\ Rev.\  D {\bf 75}, 083526 (2007)
  [arXiv:astro-ph/0611925].

\bibitem{Bruijn:2008zz}
  R.~Bruijn, PhD thesis,
  The Antares Neutrino Telescope: Performance studies and analysis of  first
  data.

\bibitem{photon}
  N.~F.~Bell and T.~D.~Jacques,
  arXiv:0811.0821 [astro-ph];
  I.~Cholis, G.~Dobler, D.~P.~Finkbeiner, L.~Goodenough and N.~Weiner,
  arXiv:0811.3641 [astro-ph];
  G.~Bertone, M.~Cirelli, A.~Strumia and M.~Taoso,
  arXiv:0811.3744 [astro-ph];
  E.~Nardi, F.~Sannino and A.~Strumia,
  arXiv:0811.4153 [hep-ph];
  J.~Zhang, X.~J.~Bi, J.~Liu, S.~M.~Liu, P.~f.~Yin, Q.~Yuan and S.~H.~Zhu,
  arXiv:0812.0522 [astro-ph].

\bibitem{Bhadra:2008cg}
  A.~Bhadra and R.~K.~Dey,
  arXiv:0812.1845 [astro-ph].

\bibitem{Hisano:2008ah}
  J.~Hisano, M.~Kawasaki, K.~Kohri and K.~Nakayama,
  arXiv:0812.0219 [hep-ph].

\bibitem{Moore:1999gc}
  B.~Moore, T.~R.~Quinn, F.~Governato, J.~Stadel and G.~Lake,
  Mon.\ Not.\ Roy.\ Astron.\ Soc.\  {\bf 310}, 1147 (1999)
  [arXiv:astro-ph/9903164].

\bibitem{eke01}
  V.~R.~Eke, J.~F.~Navarro and M.~Steinmetz,
  Astrophys.\ J.\  {\bf 554}, 114 (2001).

\bibitem{bullock01}
  J.~S.~Bullock {\it et al.},
  Mon.\ Not.\ Roy.\ Astron.\ Soc.\  {\bf 321}, 559 (2001).

\bibitem{lavalle08}
  J.~Lavalle, Q.~Yuan, D.~Maurin and X.~J.~Bi,
  Astron.\ Astrophys.\  {\bf 479}, 427 (2008).

\bibitem{sim}
  G.~Tormen, A.~Diaferio and D.~Syer, Mon.\ Not.\ Roy.\ Astron.\ Soc.\ {\bf
  299}, 728 (1998);
  A.~A.~Klypin, S.~Gottlober and A.~V.~Kravtsov,
  Astrophys.\ J.\  {\bf 516}, 530 (1999).
  [arXiv:astro-ph/9708191];
  B.~Moore, S.~Ghigna, F.~Governato, G.~Lake, T.~Quinn, J.~Stadel and P.~Tozzi,
  Astrophys.\ J.\  {\bf 524} (1999) L19;
  S.~Ghigna, B.~Moore, F.~Governato, G.~Lake, T.~Quinn and J.~Stadel,
  Astrophys.\ J.\  {\bf 544}, 616 (2000).
  [arXiv:astro-ph/9910166];
  V.~Springel, S.~D.~M.~White, G.~Tormen and G.~Kauffmann,
  Mon.\ Not.\ Roy.\ Astron.\ Soc.\  {\bf 328}, 726 (2001).
  [arXiv:astro-ph/0012055];
  A.~R.~Zentner and J.~S.~Bullock,
  Astrophys.\ J.\  {\bf 598}, 49 (2003).
  [arXiv:astro-ph/0304292];
  G.~De Lucia {\it et al.},
  Mon.\ Not.\ Roy.\ Astron.\ Soc.\  {\bf 348}, 333 (2004).
  [arXiv:astro-ph/0306205];
  J.~Diemand, B.~Moore and J.~Stadel,
  Nature {\bf 433}, 389 (2005)
  [arXiv:astro-ph/0501589].

\bibitem{Diemand:2004kx}
  J.~Diemand, B.~Moore and J.~Stadel,
  Mon.\ Not.\ Roy.\ Astron.\ Soc.\  {\bf 352}, 535 (2004)
  [arXiv:astro-ph/0402160].

\bibitem{Madau:2008fr}
  P.~Madau, J.~Diemand and M.~Kuhlen,
  arXiv:0802.2265 [astro-ph].



\end{thebibliography}
\end{document}